\documentclass[a4paper,10pt,preprint,twocolumn,3p]{elsarticle} 

\usepackage{lineno}
\usepackage{mathtools}
\usepackage{bm,amssymb}
\usepackage{subcaption}
\usepackage{booktabs}
\usepackage{color} 
\usepackage{natbib}
\usepackage{physics}
\usepackage[dvipsnames]{xcolor}
\usepackage[pdftex,colorlinks,citecolor=blue,bookmarks]{hyperref}

\journal{Physics Letters B}

\biboptions{numbers,sort&compress}
\begin{document}
\onecolumn
\begin{frontmatter}
\title{Impact of the leading-order short-range  nuclear matrix element on the neutrinoless double-beta decay of medium-mass and heavy nuclei
}%

\author{Lotta Jokiniemi}
\ead{jokiniemi@ub.edu}

\author{Pablo Soriano}
\ead{pablosoriano@ub.edu}

\author{Javier Men\'{e}ndez}
\ead{javier.menendez@fqa.ub.edu}

\address{%
Departament de Física Qu\`{a}ntica i Astrof\'{i}sica and Institut de Ci\`{e}ncies del Cosmos, Universitat de Barcelona, 08028 Barcelona, Spain
}%

\date{\today}

\begin{abstract}
We evaluate the leading-order short-range nuclear matrix element for the neutrinoless double-beta ($0\nu\beta\beta$) decay of the nuclei most relevant for experiments, including $^{76}$Ge, $^{100}$Mo, $^{130}$Te and $^{136}$Xe. In our calculations, performed with the nuclear shell model and proton-neutron quasiparticle random-phase approximation (pnQRPA) methods, we estimate the coupling of this term by the contact charge-independence-breaking coupling of various nuclear Hamiltonians. Our results suggest a significant impact of the short-range matrix element, which is about $15\%-50\%$ and $30\%-80\%$ of the standard $0\nu\beta\beta$-decay long-range matrix element for the shell model and pnQRPA, respectively. Combining the full matrix elements with the results from current $0\nu\beta\beta$-decay experiments we find that, if both matrix elements carry the same sign, these searches move notably toward probing the inverted mass ordering of neutrino masses.
\end{abstract}
\begin{keyword}Neutrinoless double-beta decay, nuclear matrix elements, quasiparticle random-phase approximation, nuclear shell model
\end{keyword}
\end{frontmatter}

\section{\label{sec:intro}Introduction}
Neutrinoless double-beta ($0\nu\beta\beta$) decay is a hypothetical nuclear process where two neutrons turn into two protons and only two electrons are emitted. Since two leptons are created, an observation of this decay would point to an event beyond the standard model (SM) of particle physics. Further, $0\nu\beta\beta$ decay can only occur if the neutrino is a Majorana particle---its own antiparticle---unlike any other fundamental particle known. The observation of $0\nu\beta\beta$ decay would therefore shed light on beyond-SM physics such as the matter-antimatter asymmetry of the universe~\cite{Fukugita:1986hr,Davidson:2008bu}, and the nature of the neutrino~\cite{Avignone2008,Vergados2012,Engel2017,Dolinski:2019nrj,Ejiri:2019ezh}.

Double-beta decay with the emission of neutrinos, which is allowed by the SM, has already been observed in about a dozen nuclei~\cite{Barabash2020} where $\beta$ decay is energetically forbidden or very suppressed. The neutrinoless mode is under massive searches by several large-scale experiments worldwide \cite{Adams2021,Agostini2020,Anton2019,Alvis2019,Azzolini2019,Gando2016}, with the most stringent half-life limits given by $t_{1/2}^{0\nu}>1.8\times 10^{26}\,$y for $^{76}$Ge~\cite{Agostini2020}, $t_{1/2}^{0\nu}>1.07\times 10^{26}\,$y for $^{136}$Xe~\cite{Gando2016} and $t_{1/2}^{0\nu}>2.2\times 10^{25}\,$y for $^{130}$Te \cite{Adams2021}. However, in order to interpret the experimental results it is crucial to have reliable $0\nu\beta\beta$-decay nuclear matrix elements (NMEs), which need to be predicted from nuclear theory. 

Even though several beyond-SM mechanisms can trigger $0\nu\beta\beta$ decay, one of the best motivated scenarios involves the exchange of the three known light neutrinos. The corresponding NMEs can be obtained with the nuclear shell model (NSM)~\cite{Menendez18,Horoi16b,Coraggio20,Iwata:2016cxn}, the proton-neutron quasiparticle random-phase approximation (pnQRPA) method~\cite{Mustonen13,Hyvarinen2015,Jokiniemi2018,Terasaki20,Simkovic18,Fang18}, energy-density functional theory~\cite{Rodriguez10,Vaquero14,Song17} or the interacting boson model~\cite{Barea15,Deppisch:2020ztt}. Ab initio methods also provide NMEs for selected $\beta\beta$ nuclei: the in-medium similarity renormalization group (IMSRG)~\cite{Yao:2019rck} and coupled cluster~\cite{Novario:2020dmr} frameworks for $^{48}$Ca, and the valence-space IMSRG for $^{48}$Ca, $^{76}$Ge and $^{82}$Se~\cite{Belley:2020ejd}.
However, present predictions for the NMEs disagree by more than a factor of two~\cite{Engel2017}.
In addition, many of these many-body methods typically overestimate matrix elements driven by the nuclear spin~\cite{Towner1987,Martinez-Pinedo:1996zvt,Caurier:2011gi,Horoi16b,KamLAND-Zen:2019imh}, a feature sometimes corrected by ``quenching'' the value of the axial-vector coupling $g_{\rm A}\simeq1.27$. The exception are ab initio methods, which reproduce $\beta$-decay matrix elements well without additional adjustments~\cite{Gysbers2019}. Nonetheless, if and to what extent $0\nu\beta\beta$-decay NMEs need to be corrected remains an open question~\cite{Menendez:2011qq,Engel2017}.
A complementary avenue is to use nuclear physics experiments to constrain the NMEs~\cite{Freeman:2012hr,Brown:2014yda,Cappuzzello:2018wek,Shimizu:2017qcy,Jokiniemi:2020ydy,Romeo:2021zrn}, but observables well correlated with $0\nu\beta\beta$ decay are also hard to access~\cite{Lenske:2019iwu,Rebeiro:2020wvo}.
In sum, NME uncertainties complicate the extraction of additional physics information from $0\nu\beta\beta$-decay experiments.

Furthermore, recently Refs.~\cite{Cirigliano2018,Cirigliano2019} introduced a previously unacknowledged short-range matrix element which appears at leading order in
light-neutrino-exchange $0\nu\beta\beta$ decay. This brings an additional, potentially significant uncertainty to $0\nu\beta\beta$-decay NMEs, especially because the value of the hadronic coupling associated with the new term is not known. First quantum Monte Carlo studies in very light $^{12}$Be, estimating the new coupling from the charge-independence-breaking (CIB) coupling of nuclear Hamiltonians, indicate that this term could amount to as much as $80\%$ of the standard long-range NME~\cite{Cirigliano2019}. The new term was also studied
in $^{48}$Ca with coupled cluster theory~\cite{Novario:2020dmr}. More recently,
Refs.~\cite{Cirigliano:2020dmx,Cirigliano:2021qko} provide synthetic data to fix the coupling of the short-range term in ab initio calculations, a procedure leading to a $43(7)\%$ enhancement of the $^{48}$Ca IMSRG NME~\cite{Wirth2021}.

In this Letter, we extend these studies and explore for the first time the short-range $0\nu\beta\beta$-decay NME in a wide range of $\beta\beta$ emitters, including all nuclei used in the most advanced experiments~\cite{Agostini2020,Alvis2019,Azzolini2019,Adams2021,Anton2019,Gando2016} and proposals for next generation searches~\cite{Abgrall:2017syy,Kharusi:2018eqi,CUPIDInterestGroup:2019inu}. We perform many-body calculations of medium-mass and heavy nuclei with nucleon number $A=48-136$ with the pnQRPA and large-scale NSM frameworks commonly used to obtain long-range NMEs, and we take CIB couplings to estimate the size of the short-range NMEs. Finally, we analyze the impact of this new term on current $0\nu\beta\beta$-decay searches combining in a consistent manner the likelihood functions of the most constraining experiments with the full NMEs for different nuclei.

\section{\label{sec:doublebeta}Neutrinoless Double-Beta Decay}

The $0\nu\beta\beta$-decay half-life can be written as~\cite{Engel2017,Cirigliano2018}
\begin{equation}
[t_{1/2}^{0\nu}]^{-1}=G_{0\nu}\,g_{\rm A}^4\,|M_{\rm L}^{0\nu}+M_{\rm S}^{0\nu}|^2\,\frac{m^2_{\beta\beta}}{m_e^2}\;,
  \label{eq:half-life}
\end{equation}
where $G_{0\nu}$ is a phase-space factor for the final-state leptons~\cite{Kotila2012}, 
and $M_{\rm L}^{0\nu}$ and $M_{\rm S}^{0\nu}$ are the long- and short-range NMEs, with unknown relative sign. The effective Majorana mass $m_{\beta\beta}=\sum_i U_{ei}m_i$ (normalized to the electron mass $m_e$) characterizes the lepton-number violation and depends on the neutrino masses $m_i$ and mixing matrix $U$.

The matrix element $M_{\rm L}^{0\nu}$ denotes the standard light-neutrino-exchange matrix element, which can be written in the familiar way~\cite{Engel2017}
\begin{equation}
M_{\rm L}^{0\nu}=M_{\rm GT}^{0\nu}-\Big(\frac{g_{\rm V}}{g_{\rm A}}\Big)^2M_{\rm F}^{0\nu}+M_{\rm T}^{0\nu}\;,
\label{eq:m_0v}
\end{equation}
with Gamow-Teller, Fermi and tensor contributions $M_{\rm GT}^{0\nu}$, $M_{\rm F}^{0\nu}$ and $M_{\rm T}^{0\nu}$, and vector coupling $g_{\rm V}=1.0$. The calculation of the matrix elements involves, in addition to the initial and final states $0^+_i$ and $0^+_f$, a sum over intermediate states, carried out explicitly in the pnQRPA~\cite{Hyvarinen2015}. Alternatively, for our NSM results we use the closure approximation to perform this sum analytically, so that the dominant Gamow-Teller term reads
\begin{align}
\label{eq:Mgt}
  & M^{0\nu}_\text{GT} = \frac{2R}{\pi g_{\rm A}^2} \times \\
&\langle 0^+_f|
\sum_{m,n}
\tau^-_m\tau^-_n\,{\bm \sigma}_m{\bm \sigma}_n
\int \frac{j_0(qr)h_\text{GT}(q^2)\,q}{q+E} {\rm d}q| 0^+_i\rangle, \nonumber
\end{align}
with sum over the spin $\bm{\sigma}$ and isospin $\tau^-$ operators of all $A$ nucleons, momentum transfer $q$, a Bessel function $j_0$, and $R=1.2 A^{1/3}\,$fm. We use as average energy for the intermediate states $E=0$. 
The matrix element also depends on a neutrino potential, with $h_\text{GT}(0)=g_{\rm A}^2$ and additional $q$-dependent subleading terms, regularized with a dipole as in previous pnQRPA~\cite{Hyvarinen2015,Simkovic2008} and NSM~\cite{Menendez18} studies.
The Fermi and tensor parts
follow similar expressions to~Eq.~(\ref{eq:Mgt})~\cite{Engel2017}.
Finally, we correct our many-body states with two-nucleon short-range correlations (SRCs) following the so-called CD-Bonn and Argonne parametrizations~\cite{Simkovic2009}.

\begin{table}[b]
    \caption{Couplings ($g_{\nu}^{\rm NN}$) and scales ($\Lambda$) of the Gaussian regulator considered for the short-range NME $M^{0\nu}_{S}$.}
    \centering
    \begin{tabular}{ccc}
    \toprule
    $g_{\nu}^{\rm NN}(\rm fm^2)$ &$\Lambda$ (MeV) &Ref.\\
    \midrule
    -0.67 &450 &\cite{Reinert2018}\\
    -1.01 &550 &\cite{Reinert2018}\\
    -1.44 &465 &\cite{Piarulli2016}\\
    -0.91 &465 &\cite{Piarulli2016}\\
    -1.44 &349 &\cite{Piarulli2016}\\
    -1.03 &349 &\cite{Piarulli2016}\\
     \bottomrule
    \end{tabular}
    \label{tab:regulators}
\end{table}

The short-range matrix element 
connects directly the initial and final nuclei~\cite{Cirigliano2018}
\begin{align}\label{eq:contact}
  M^{0\nu}_{\rm S} = \frac{2R}{\pi g_{\rm A}^2}
\langle 0^+_f|
\sum_{m,n}
\tau^-_m\tau^-_n\,\!\!
\int \!j_0(qr)h_{\rm S}(q^2)\,q^2 {\rm d}q| 0^+_i\rangle,
\end{align}
where we choose to regularize the contact term with a Gaussian in
the neutrino potential:
\begin{equation}
h_{\rm S}(q^2)=2g_{\nu}^{\rm NN}\,e^{-q^2/(2\Lambda^2)}\;,
\label{eq:h-contact}
\end{equation}
with $\Lambda$ the scale of the regulator. 


The coupling $g_{\nu}^{\rm NN}$, not part of the SRCs, can only be fixed by fitting to lepton-number-violating data---currently unavailable---or synthetic data~\cite{Cirigliano:2020dmx,Cirigliano:2021qko}---only accessible to ab initio calculations. Here we follow Ref.~\cite{Cirigliano2019} and estimate its value by considering the CIB term of different nuclear Hamiltonians, restricted to cases with a Gaussian regulator.
This strategy carries some uncertainty since two low-energy constants are needed to fix $g_{\nu}^{\rm NN}$, while only one of them can be extracted from the CIB term---the other one is assumed to have the same value. Nonetheless, the empirical CIB was well reproduced in Refs.~\cite{Cirigliano:2020dmx,Cirigliano:2021qko} with the same strategy used to obtain their synthetic lepton-number-violating data.
Table~\ref{tab:regulators} shows the $g_{\nu}^{\rm NN}-\Lambda$ pairs considered in our work.

\section{\label{sec:many-body}Many-Body NME Calculations}

We perform NSM calculations with the coupled code NATHAN~\cite{Caurier2005}.
We use the KB3G~\cite{Poves:2000nw} interaction in the $pf$-shell ($0f_{7/2}$, $1p_{3/2}$, $0f_{5/2}$ and $1p_{1/2}$ orbitals) for $A=48$, the RG.5-45~\cite{Menendez2009} interaction with $1p_{3/2}$, $0f_{5/2}$, $1p_{1/2}$ and $0g_{9/2}$ orbitals
for $A=76, 82$ and the GCN5082~\cite{Menendez2009} interaction with $0g_{7/2}$, $1d_{5/2}$, $1d_{5/2}$, $2s_{1/2}$ and $0h_{11/2}$ orbitals for $A=124-136$. In all cases our valence space is common to protons and neutrons. Overall, with the NSM we study seven $0\nu\beta\beta$ decays: $^{48}$Ca, $^{76}$Ge, $^{82}$Se, $^{124}$Sn, $^{128,130}$Te and $^{136}$Xe. 
We consider all nuclear configurations in the full valence space except in $^{124}$Te (final nucleus of the $^{124}$Sn $\beta\beta$ decay)
which is limited to seniority $v\leq5$ states (up to five broken zero-angular-momentum pairs) instead of the full $v\leq6$ space. We have checked that the corresponding NMEs are converged to the percent level. Compared to the pnQRPA calculations described below, the decays of $^{96}$Zr, $^{100}$Mo and $^{116}$Cd are still out of reach for the NSM.

On the other hand, we use the spherical pnQRPA method as in Refs.~\cite{Jokiniemi2018,Jokiniemi2019}. The large no-core single-particle bases consist of 18 orbitals for $A=76,82$ nuclei, 25 orbitals for $A=96,100$, and 26 orbitals for $A=124-136$. In all, in this framework we study the decays of $^{76}$Ge, $^{82}$Se, $^{96}$Zr, $^{100}$Mo, $^{116}$Cd, $^{124}$Sn, $^{128,130}$Te and $^{136}$Xe, excluding $^{48}$Ca because the pnQRPA does not describe doubly-magic nuclei reliably.
We take the single-particle energies 
from a Coulomb-corrected Woods-Saxon potential optimized for nuclei close to the $\beta$-stability line~\cite{Bohr1969},
but in the vicinity of the Fermi surface we slightly modify them to better reproduce the low-lying spectra of neighboring odd-mass nuclei.
The quasiparticle spectra, needed in the pnQRPA diagonalization, follow the solution of the BCS equations for protons and neutrons. We use the two-body interaction derived from the Bonn-A
potential \cite{Holinde1981}, fine-tuning the proton and neutron pairing parameters to reproduce the phenomenological pairing gaps. The residual Hamiltonian for the pnQRPA calculation contains two adjustable scaling factors: the particle-hole $g_{\rm ph}$  and particle-particle $g_{\rm pp}$ parameters.
We fix $g_{\rm ph}$ to reproduce the centroid of the Gamow-Teller giant resonance, 
and $g_{\rm pp}$ to the two-neutrino $\beta\beta$-decay half-life according to the partial isospin-symmetry restoration scheme introduced in Ref.~\cite{Simkovic2013}.

\section{\label{sec:results}Results and Discussion}

\begin{table*}
\centering
\caption{Long- and short-range $0\nu\beta\beta$-decay matrix elements $M^{0\nu}_{\rm L}$ and $M^{0\nu}_{\rm S}$ calculated with the pnQRPA and nuclear shell model (NSM) for several nuclei. The ranges cover results for neutrino potentials with the couplings and regulators in Table~\ref{tab:regulators}, combined with Argonne and CD-Bonn short-range correlations.}
\makebox[\linewidth]{
\begin{tabular}{lcccccc}
\toprule
& \multicolumn{3}{c}{pnQRPA} &\multicolumn{3}{c}{NSM}\\
\cmidrule{2-4} \cmidrule{5-7}
Nucleus & $M_{\rm L}^{0\nu}$ &$M_{\rm S}^{0\nu}$& $M_{\rm S}^{0\nu}/M_{\rm L}^{0\nu}(\%)$ & $M_{\rm L}^{0\nu}$ &$M_{\rm S}^{0\nu}$ &$M_{\rm S}^{0\nu}/M_{\rm L}^{0\nu}(\%)$\\
\midrule
$^{48}$Ca  &              &              &          &$0.96 - 1.05$ &$0.22 - 0.65$ &$23 - 62$\\
$^{76}$Ge  &$4.72 - 5.22$ &$1.49 - 3.80$ &$32 - 73$ &$3.34 - 3.54$ &$0.52 - 1.49$ &$15 - 42$\\
$^{82}$Se  &$4.20 - 4.61$ &$1.27 - 3.24$ &$30 - 70$ &$3.20 - 3.38$ &$0.48 - 1.38$ &$15 - 41$\\
$^{96}$Zr  &$4.22 - 4.63$ &$1.24 - 3.19$ &$29 - 69$\\
$^{100}$Mo &$3.40 - 3.95$ &$1.66 - 4.26$ &$49 - 108$\\
$^{116}$Cd &$4.24 - 4.57$ &$1.10 - 2.80$ &$26 - 61$\\
$^{124}$Sn &$4.72 - 5.29$ &$1.69 - 4.28$ &$36 - 81$ &$3.20 - 3.41$ &$0.54 - 1.58$ &$17 - 46$\\
$^{128}$Te &$3.92 - 4.50$ &$1.37 - 3.45$ &$35 - 77$ &$3.56 - 3.80$ &$0.61 - 1.76$ &$17 - 46$\\
$^{130}$Te &$3.46 - 3.89$ &$1.18 - 3.05$ &$34 - 77$ &$3.26 - 3.48$ &$0.57 - 1.64$ &$17 - 47$\\
$^{136}$Xe &$2.53 - 2.80$ &$0.76 - 1.95$ &$30 - 70$ &$2.62 - 2.79$ &$0.45 - 1.31$ &$17 - 47$\\
\bottomrule
\end{tabular}
}
\label{tab:NMEs}
\end{table*}

We calculate the $0\nu\beta\beta$-decay short- and long-range NMEs for ten heavy nuclei, listed in Table~\ref{tab:NMEs}. For both NSM and pnQRPA, in all transitions the standard matrix element $M_{\rm L}^{0\nu}$ is larger than the new term  $M_{\rm S}^{0\nu}$. Nonetheless, Table \ref{tab:NMEs} shows that in both many-body frameworks the contribution of the short-range matrix element is significant: in the pnQRPA the ratios of the short- over long-range NMEs typically range between $30\%-80\%$; in the NSM the ratios are slightly more moderate, between $15\%-50\%$. Within a given method, the relative size of $M_{\rm S}^{0\nu}$ is in general rather stable.
Our results indicate that the short-range contribution can considerably impact the expected rates of current and future  $0\nu\beta\beta$-decay experiments. Therefore, the new term should be calculated in heavy nuclei using more consistent $g_{\nu}^{\rm NN}$ values.

The NME ranges in Table~\ref{tab:NMEs} are much wider in the case of the short-range term than for the standard matrix element, as the difference between the lower and upper $M_{\rm S}^{0\nu}$ values can be up to a factor of three for both methods. This partially reflects the variety of couplings $g_{\nu}^{\rm NN}$ and regulator scales $\Lambda$ in Table~\ref{tab:regulators}: the smallest short-range values are always given by $g_{\nu}^{\rm NN}=-0.67~{\rm fm}^2$ and $\Lambda=450$ MeV (with Argonne SRCs), while the largest ones involve in all cases $g_{\nu}^{\rm NN}=-1.44~{\rm fm}^2$ and $\Lambda=465$ MeV (with CD-Bonn SRCs). In contrast, the small differences in the long-range $M_{\rm L}^{0\nu}$ are driven by the SRCs, with the lower values corresponding to Argonne SRCs and the upper ones to CD-Bonn SRCs. 

\begin{figure}
\centering
\includegraphics[width=\linewidth]{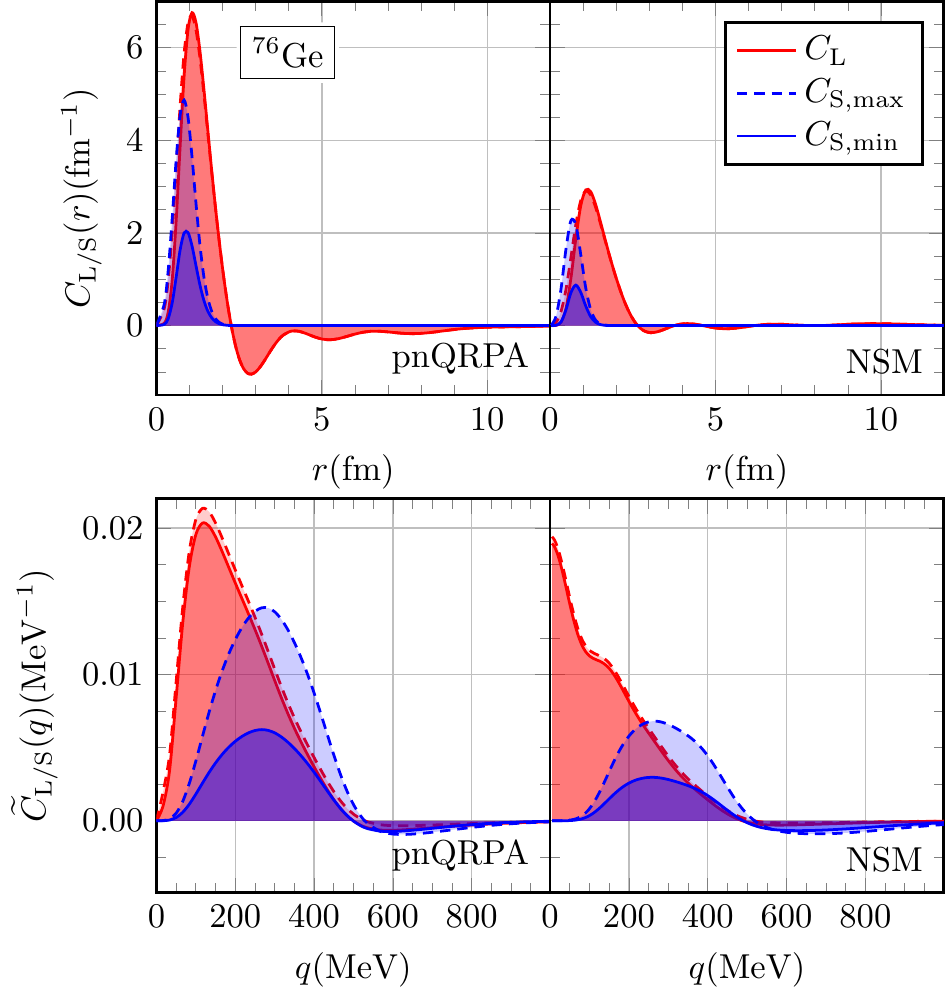}
\caption{Long- (red) and short-range (blue) $^{76}$Ge radial (top) and momentum (bottom panels) matrix-element distributions for the pnQRPA (left) and NSM (right panels). Solid and dashed lines indicate Argonne and CD-Bonn short-range correlations, respectively. Short-range matrix-element distributions are shown for the neutrino potential parameters leading to extreme results: $g_{\nu}^{\rm NN}=-1.44~{\rm fm}^2$ with regulator $\Lambda=465~{\rm MeV}$ (light blue), and $g_{\nu}^{\rm NN}=-0.67~{\rm fm}^2$ with $\Lambda=450~{\rm MeV}$ (dark blue).
}
\label{fig:76Ge}
\end{figure}

In order to study the short- and long-range NMEs in more detail, Fig.~\ref{fig:76Ge} shows their radial and momentum distributions, denoted by $C_{\rm L/S}(r)$ and $\widetilde{C}_{\rm L/S}(q)$, for $^{76}$Ge. The distributions satisfy 
\begin{equation}
\int C_{\rm L/S}(r){\rm d}r=M^{0\nu}_{\rm L/S}=\int \widetilde{C}_{\rm L/S}(q){\rm d}q\;,   
\end{equation}
where $r=|r_{n}-r_{m}|$ is the distance between the two decaying nucleons. Figure~\ref{fig:76Ge} shows the short-range distributions calculated with the two extreme neutrino potentials: $g_{\nu}^{\rm NN}=-0.67~{\rm fm}^2$, $\Lambda=450$~MeV with Argonne SRCs (dark blue area), and $g_{\nu}^{\rm NN}=-1.44~{\rm fm}^2$, $\Lambda=465$~MeV with CD-Bonn SRCs (light blue). While the shapes of the two blue areas are similar, the size of the light one is clearly larger in all cases. For the long-range term, Fig.~\ref{fig:76Ge} shows results for Argonne (solid red curve) and CD-Bonn (dashed red) SRCs, hardly distinguishable in both NSM and pnQRPA. As expected, the radial distribution of $M^{0\nu}_{\rm S}$ involves shorter internucleon distances than the one of the long-range $M^{0\nu}_{\rm L}$, and its momentum distribution reaches larger momentum transfers. Apart from the consistently smaller NME values obtained with the NSM, the overall behavior of the matrix-element distributions is quite similar in both frameworks.

Figure~\ref{fig:76Ge} also shows differences between many-body methods. In the pnQRPA, the radial distribution of the long-range NME gets a sizeable cancellation from distances $r\gtrsim2.5$~fm, which is much milder in the NSM. This rather well-known feature of the pnQRPA~\cite{Simkovic2008,Simkovic2009,Hyvarinen2015} partly explains the relatively larger size of $M_{\rm S}^{0\nu}$ with respect to $M_{\rm L}^{0\nu}$, since there are no cancellations in the short-range NME radial distribution.
Furthermore, Fig.~\ref{fig:76Ge} also highlights that the short-range pnQRPA NME extends to longer distances than the NSM one, whereas the positive contribution to the  pnQRPA long-range NME is concentrated at shorter distances. This behavior also leads to larger pnQRPA $M_{\rm S}^{0\nu}/M_{\rm L}^{0\nu}$ ratios. In momentum space, the pnQRPA long-range  NME distribution reaches larger momentum transfers, while the NSM one does not vanish at $q=0$ because of our closure energy $E=0$.

\begin{figure}[t]
\centering
\includegraphics[width=\linewidth]{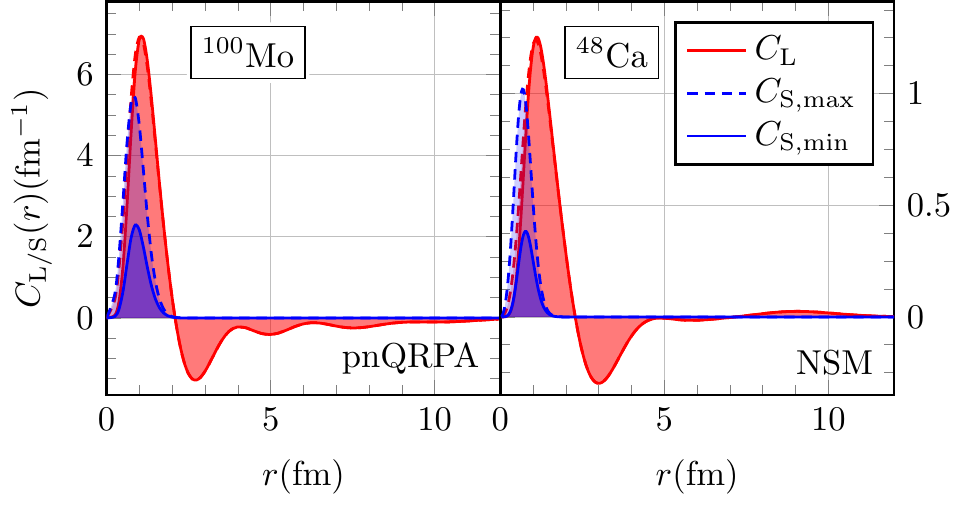}
\caption{Radial long- and short-range matrix-element distributions for {$^{100}$Mo} calculated with the pnQRPA framework (left), and {$^{48}$Ca} obtained with the NSM (right panel, with scale at the right $y$ axis). Line and color codes as in Fig.~\ref{fig:76Ge}.}
\label{fig:48Ca-100Mo}
\end{figure}

Two transitions stand out with the largest relative short-range $M^{0\nu}_{\rm S}$ values:  $^{100}$Mo for the pnQRPA, and $^{48}$Ca for the NSM. Figure~\ref{fig:48Ca-100Mo} shows the radial short- and long-range NME distributions for these two cases. Apart from the different scales,
the $^{48}$Ca radial distribution resembles the $^{76}$Ge pnQRPA long-range one in Fig.~\ref{fig:76Ge}: there is a sizeable cancellation in $C_{\rm L}(r)$ at distances $r\approx(2-5)$~fm, not observed in any other NSM decay.
Such cancellation never occurs for the short-range $C_{\rm S}(r)$, which explains the larger relative contribution of $M^{0\nu}_{\rm S}$ for this nucleus.
The relative size of our $^{48}$Ca short-range NME is similar to the ab initio result from Ref.~\cite{Wirth2021}, however obtained with a different coupling and regulator scheme.
Figure~\ref{fig:48Ca-100Mo} also shows a more marked cancellation in the pnQRPA long-range $^{100}$Mo NME than in $^{76}$Ge. This exceptionally large cancellation, not present in any other nucleus, is explained by a notable negative contribution at low momenta which reduces the value of $M_{\rm L}^{0\nu}$. This behavior is driven by the $1^+$ multipole which dominates at low-$q$ values, as observed in previous pnQRPA works~\cite{Simkovic2008,Hyvarinen2015}. A similar feature appears in light nuclei studied with quantum Monte Carlo~\cite{Pastore2018} and the NSM.

\begin{figure}[t]
\centering
\includegraphics[width=8cm]{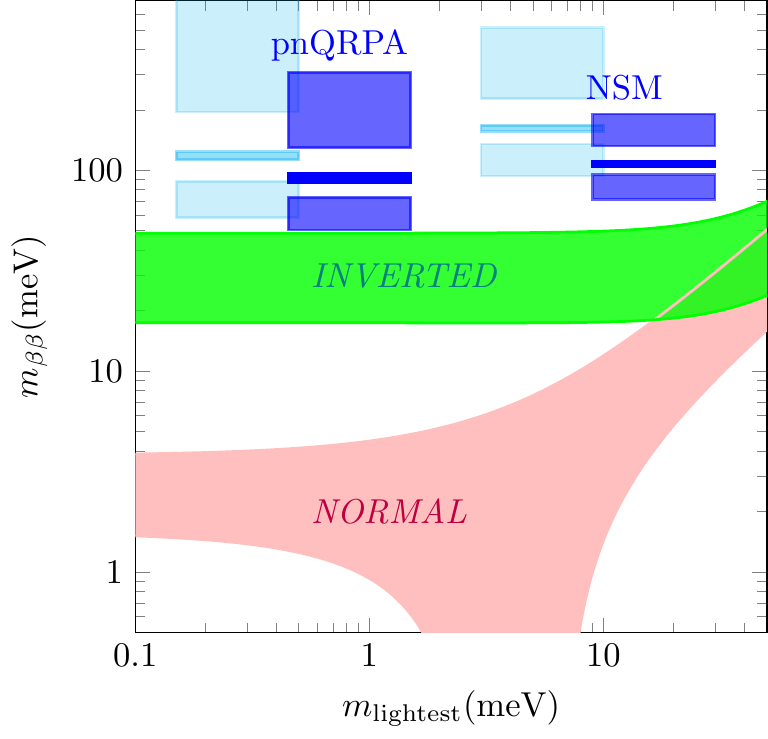}%
\caption{Effective Majorana mass $ m_{\beta\beta}$ in terms 
of the lightest neutrino mass $m_{\rm lightest}$ assuming the normal (pink) or inverted (green) ordering of neutrino masses \cite{Salas2021,Dell'Oro2016}, compared to the exclusion (blue) bands which combine data \cite{Biller2021} from $0\nu\beta\beta$-decay experiments~\cite{Adams2021,Agostini2020,Anton2019,Gando2016} and pnQRPA or NSM NMEs. The upper, middle and lower exclusion bands correspond to NME ranges for $M_{\rm L}^{0\nu}-M_{\rm S}^{0\nu}$, $M_{\rm L}^{0\nu}$ and $M_{\rm L}^{0\nu}+M_{\rm S}^{0\nu}$, accordingly.
The cyan bands correspond to a ``quenching'' scenario, see text for details.
}
\label{fig:masses}
\end{figure}

The $M_\text{L}^{0\nu}$ matrix elements in Table~\ref{tab:NMEs} assume $g_{\rm A}=1.27$. Related NSM and pnQRPA $\beta$ and two-neutrino $\beta\beta$ decay rates obtained this way are known to be overestimated, calling for corrections usually known as ``$g_A$ quenching''. While the implications to $0\nu\beta\beta$-decay NMEs are not clear~\cite{Engel2017}, they would only affect the long-range NME, leading to a larger relative impact of the short-range term. We consider such ``quenching'' scenario to provide more conservative estimates. On the one hand, we calculate pnQRPA $(g_{\rm A}^{\rm eff}/g_{\rm A})^2M_\text{L}^{0\nu}$ values with $g_\text{pp}$ fitted to reproduce two-neutrino $\beta\beta$-decay half-lifes with $g_A^\text{eff}=1$. The results are reduced by about 20\%. In a similar spirit, we multiply the NSM $M_{\rm GT}^{0\nu}$ terms by $1/g_{\rm A}^2\simeq 0.6$, which reduces $M_\text{L}^{0\nu}$ by 30\% or so. For $^{48}$Ca, this brings the NSM in good agreement with ab initio theory~\cite{Belley:2020ejd,Yao:2019rck,Wirth2021}. With these rough estimates, the short-range NME contribution increases by about $25\%$ in the pnQRPA, reaching about $40\%-100\%$ in $^{76}$Ge, $^{130}$Te and $^{136}$Xe. For the NSM, the impact of the short-range term would be enhanced by about $50\%$, typically up to $25\%-70\%$. 

Finally, we explore the impact of our $M_{\rm S}^{0\nu}$ results on the current reach of the experimental $0\nu\beta\beta$-decay program in terms of bounds on $m_{\beta\beta}$. In order to obtain stronger limits, we follow Ref.~\cite{Biller2021} to combine our $^{76}$Ge, $^{130}$Te and $^{136}$Xe NMEs
with the parameterized likelihood functions of the $0\nu\beta\beta$-decay rate, extracted from the CUORE \cite{Adams2021}, GERDA (Phase II) \cite{Agostini2020},  EXO-200 \cite{Anton2019} and KamLAND-Zen \cite{Gando2016} experiments. To combine the likelihood functions, we convert decay rates into effective Majorana masses according to Eq. \eqref{eq:half-life} with our NMEs and the phase-space factors of Ref. \cite{Kotila2012}. This way, we obtain 90\% confidence level (CL) upper bounds on $m_{\beta\beta}$ from the 90\% CL upper bound of the combined likelihood function (related to the Bayesian rather than the frequentist limit \cite{Biller2021}).

We consider three different scenarios to derive bounds on $m_{\beta\beta}$: a baseline using the standard $M^{0\nu}_{\rm L}$; an optimistic scenario assuming common signs for the short- and long-range NMEs  (with $M^{0\nu}_{\rm L}+M^{0\nu}_{\rm S}$ in Table~\ref{tab:NMEs}); and a pessimistic one where the short-range part cancels the standard matrix element (with $M^{0\nu}_{\rm L}-M^{0\nu}_{\rm S}$). The first consistent determination of the short-range matrix element in $^{48}$Ca supports the optimistic scenario~\cite{Wirth2021}.
We take a matrix-element uncertainty given by the extreme values of $M^{0\nu}_{\rm L}$ and $M^{0\nu}_{\rm L}\pm M^{0\nu}_{\rm S}$ obtained from the set of 12 calculations corresponding to the six $g_{\nu}^{\rm NN}-\Lambda$ pairs in Table~\ref{tab:regulators} and two SRCs. For $^{76}{\rm Ge}$, $^{130}{\rm Te}$ and $^{136}{\rm Xe}$, the extreme values are always given by the same $g_{\nu}^{\rm NN}-{\rm SRC}$ combinations.

Figure \ref{fig:masses} compares the constraints on $m_{\beta\beta}$ in the three scenarios with the bands corresponding to the normal and inverted neutrino-mass orderings, obtained with neutrino-oscillation data~\cite{Salas2021} as described in Ref.~\cite{Dell'Oro2016}. The widths of the blue bands in Fig.~\ref{fig:masses} correspond to the ranges of $M^{0\nu}_{\rm L}$ (middle), $M^{0\nu}_{\rm L}-M^{0\nu}_{\rm S}$ (top) and $M^{0\nu}_{\rm L}+M^{0\nu}_{\rm S}$ (bottom) in Table~\ref{tab:NMEs}, and are much larger once the new short-range term is included. The pnQRPA bands are dominated by the likelihood function of GERDA partly due to the large $^{76}$Ge NME, while the next-constraining experiments are KamLAND-Zen, EXO-200 and CUORE, in that order. In the NSM the hierarchy is similar. In the scenario that both matrix elements carry the same sign, our results indicate that the reach of current experiments approaches the inverted mass-ordering region notably. This feature is more marked when using the pnQRPA NMEs, and may be more moderate if our results obtained with $g_{\rm A}=1.27$ somewhat underestimate the decay half-lives. A more conservative ``quenching'' scenario is shown by the cyan bands in Fig.~\ref{fig:masses}. On the other hand, if the signs of the two matrix elements are opposite, experiments would still be far from exploring  $m_{\beta\beta}$ values corresponding to the inverted mass ordering. In this case, the pnQRPA and NSM NMEs would be similar within uncertainties.

\section{\label{sec:conclusions}Summary}

We have calculated for the first time the short-range NME which contributes at leading order to the $0\nu\beta\beta$ decay of
medium-mass and heavy nuclei including $^{76}$Ge, $^{100}$Mo, $^{130}$Te and $^{136}$Xe.
Since the value of the coupling $g_{\nu}^{\rm NN}$ of this short-range term is not known, we estimate it by a set of CIB couplings of different Hamiltonians, together with the corresponding regulators.
We find that the new short-range NME values are a significant fraction of the standard long-range ones: typically between $30\%-80\%$ in the pnQRPA and $15\%-50\%$ in the NSM. These ranges are driven by the different couplings $g_{\nu}^{\rm NN}$ considered, and are rather stable among all nuclei. The only exceptions are $^{100}$Mo and $^{48}$Ca where the ratios are notably larger due to cancellations in the standard long-range NME. Since these cancellations are typically larger in the pnQRPA than in the NSM, for the former the relative impact of the short-range matrix element is also larger.

The new short-range term can also affect the interpretation of present and future $0\nu\beta\beta$-decay searches. To this end, we derive constraints on $m_{\beta\beta}$ using our pnQRPA and NSM NMEs to combine the likelihood functions of the most constraining experiments. We observe that if the long- and short-range NMEs carry the same sign, as suggested by a recent determination \cite{Wirth2021}, the $m_{\beta\beta}$ values
constrained by these searches clearly approach the inverted neutrino-mass region.

\section*{Acknowledgements}
We would like to thank M. Agostini, G. Benato, J. de Vries, J. Detwiler, E. Mereghetti and J. Suhonen for insightful discussions. 
This work was supported by the Finnish Cultural Foundation grant No. 00210067, the Spanish MICINN through the ``Ram\'on y Cajal'' program with grant RYC-2017-22781, the AEI ``Unit of Excellence Mar\'ia de Maeztu 2020-2023'' award CEX2019-000918-M and the AEI grant FIS2017-87534-P.


\bibliography{contact-refs}

\end{document}